\documentclass[conference,a4paper]{IEEEtran}
\IEEEoverridecommandlockouts
\usepackage{cite}
\usepackage{amsmath,amssymb,amsfonts}
\usepackage{algorithmic}
\usepackage{graphicx}
\usepackage{textcomp}
\usepackage{xcolor}
\def\BibTeX{{\rm B\kern-.05em{\sc i\kern-.025em b}\kern-.08em
    T\kern-.1667em\lower.7ex\hbox{E}\kern-.125emX}}
\usepackage{multirow}
\usepackage{url}
\usepackage{listings}
\usepackage{color}
\usepackage{booktabs}
\begin{document}
\bstctlcite{IEEEexample:BSTcontrol}

\title{PAMS: Platform for Artificial Market Simulations}

\author{\IEEEauthorblockN{Masanori HIRANO}
  \IEEEauthorblockA{\textit{School of Engineering} \\
    \textit{The University of Tokyo}\\
    Tokyo, Japan \\
    research@mhirano.jp}
  \and
  \IEEEauthorblockN{Ryosuke TAKATA}
  \IEEEauthorblockA{\textit{Graduate School of Arts and Sciences} \\
    \textit{The University of Tokyo}\\
    Tokyo, Japan \\
    takata@sacral.c.u-tokyo.ac.jp}
  \and
  \IEEEauthorblockN{Kiyoshi IZUMI}
  \IEEEauthorblockA{\textit{School of Engineering} \\
    \textit{The University of Tokyo}\\
    Tokyo, Japan \\
    izumi@sys.t.u-tokyo.ac.jp}
}

\maketitle

\begin{abstract}
  This paper presents a new artificial market simulation platform, PAMS: Platform for Artificial Market Simulations.
  PAMS is developed as a Python-based simulator that is easily integrated with deep learning and enabling various simulation that requires easy users' modification.
  In this paper, we demonstrate PAMS effectiveness through a study using agents predicting future prices by deep learning.
\end{abstract}


\section{Introduction}
Artificial market simulation, one of multi-agent simulations for financial markets, is a very useful tool.
Financial markets have high complexity that comes from interaction between traders and markets.
Additionally, there is no dominant equation for phenomena happening in financial markets.
Therefore, multi-agent simulations where agents can interact with others are necessary to reproduce well-known phenomena in financial markets.
For example, \cite{Lux1999} showed that, in financial market simulations, interactions between agents (traders) are necessary to replicate common phenomena in financial markets.

Moreover, multi-agent simulations are useful tools for financial markets to test scenarios that cannot occur in real markets.
Financial market systems and regulations are crucial for market growth and stability.
For example, excessive regulation may cause side effects.
However, for discussing new regulations and systems for financial markets, the situations have not existed before.
Thus, it is difficult to estimate how the side effects appear because of the new regulation.
It means that multi-agent simulations can be an effective means to consider new situations in financial markets.

Because of those reasons, the foundation of multi-agent simulations in financial markets is essential but still has limitations and challenges.
Recently, many data analysis methods using deep learning and machine learning have appeared.
For example, in tasks such as stock price prediction, deep learning and machine learning are widely used in some researched \cite{Kim2003,Kercheval2015,Sirignano2019,Tsantekidis2017,Dixon2019,Zhang2017}.

It's not hard to imagine that such technologies are used in traders' decision-making in actual markets.

Additionally, methods combining artificial market simulation and deep learning technology have also started to be proposed.
For example, deep reinforcement learning strategies for traders are trained in artificial market simulations\cite{Maeda2020}, agent models built with data mining are used in artificial markets\cite{Hirano2022d,Hirano2022e}, and evaluating the validity of artificial market simulations using data mining techniques is also poposed\cite{Hirano2022-ica}.

Given these recent trends in the application of deep learning technology, it is important to have a foundation for the artificial market simulation platform itself to work smoothly with deep learning technologies, but this has not been realized.
As an artificial market simulation platform there are some platforms, such as "Platform for large-scale and high-frequency artificial market" (Plham) \cite{Torii2017,Plham}, PlhamJ \cite{Torii2019}, U-MART \cite{sato2001u}, Santa Fe Artificial Stock Market \cite{arthur2018asset}, and Agent-based Interactive Discrete Event Simulation (ABIDES) \cite{Byrd2020}.
In implementing artificial market simulations, compiler languages such as C++ and Java are often used from the perspective of calculation speed.
However, deep learning and machine learning libraries are often implemented in scripting languages like Python.
Therefore, combining deep learning and artificial market simulations requires calling deep learning scripts subprocesses from the artificial market simulation side or vice versa, which is not easy to implement.
Additionally, to effectively utilize artificial market simulations, it should be easy to implement new types of trader agents.
However, only Plham and PlhamJ are fully equipped with this feature.

Taking these requirements together, there is a need for an artificial market simulation platform that is composed in Python.
Python-based implementation is convenient for integration with deep learning and a class design that allows users to easily change agent code, but runtime speed is crucial, which should be addressed.
No such artificial market simulation exists yet.

Therefore, this study proposes PAMS: Platform for Artificial Market Simulations, an artificial market simulation platform that is based on Python and inherits the design philosophy of Plham.
This platform can be installed through the Python package repository PyPI\footnote{\url{https://pypi.org/project/pams/}} via commands such as pip.
Its simulation code can be easily modified.
It is also easy to integrate with deep learning.
Its performance is almost the same as that of previous platforms like Plham.
The code is publicly available at \url{https://github.com/masanorihirano/pams}, and the documentation is also available at \url{https://pams.hirano.dev/}.
In addition, along with the code, Python Jupyter Notebook implementations are also available as examples \footnote{\url{https://github.com/masanorihirano/pams/tree/main/examples}}.
You can easily try examples of PAMS using it.

In this study, we also introduce experiments and analyses of artificial market simulations implemented with deep learning.

\section{Related Works}
Multi-agent simulation has been used to analyze and understand many social phenomena since around the 1970s.
Schelling\cite{schelling1969,schelling1969} used multi-agent simulation to simulate segregation and clarified the mechanism.
Axelrod\cite{axelrod1980a,axelrod1980b} conducted a prisoner's dilemma contest in a multi-agent environment and analyzed various types of agents.
Epstein et al.\cite{epstein1996} demonstrated the potential to build artificial societies by simulating an artificial world modeled on ants and food.
Lux et al.\cite{Lux1999} pointed out the necessity of multi-agent simulation by demonstrating that the well-known phenomena observed in financial markets cannot be reproduced without the interaction between agents in simulations.
Various simulations are also being used for societal issues.
Sajjad et al.\cite{Sajjad2016} constructed a simulation of population dynamics based on real data.
Nonaka et al.\cite{Nonaka2013} constructed a people flow simulation and conducted an evacuation simulation.
Braun-Munzinger et al.\cite{Braun-Munzinger2016} constructed a multi-agent simulation for the bond market.
Kurahashi et al.\cite{Kurahashi2020} used multi-agent simulation to discuss measures to prevent the spread of COVID-19.

In social phenomena, it is difficult to analyze because there is no dominant equation or it has not been elucidated.
Therefore, simulation is useful in the field of social science\cite{Edmonds2005}.
In particular, there are studies arguing that multi-agent simulation is important\cite{Farmer2009,Battiston2016}.

While many applications are being made, there is also a discussion about the construction theory of multi-agent simulation.
Axelrod \cite{axelrod1997} advocated the "Keep It Simple Stupid" (KISS) principle in multi-agent simulation.
It stated that reproducing complex phenomena with simpler models can contribute to understanding the essence of the phenomena.
Further, Edmonds et al.\cite{edmonds2004} proposed the "Keep it Descriptive Stupid" (KIDS) principle.
It means that explainability is only a necessity, and KISS is too much for multi-agent simulations to realize various simulations.
Terano\cite{Terano2003} also discusses the possibility of agent simulation methods that go beyond the KISS principle.

This paper targets artificial market simulation, which is a multi-agent simulation particularly focused on financial markets.
The importance of social simulation and the importance of simulation in financial markets have been widely discussed\cite{Farmer2009,Battiston2016}.
As mentioned earlier, Lux et al.\cite{Lux1999} have shown the necessity of agent interaction in financial market simulation, and the necessity of multi-agent simulation has become clear.
Furthermore, Mizuta\cite{Mizuta2019} argues that multi-agent simulation in finance can contribute to financial regulation and system design.
Furthermore, the limitations of existing financial approaches are being argued by financial dignitaries, accelerating research on artificial market simulation.
The 2007–2008 financial crisis caused the collapse of investment banks and a global financial market network due to the default of housing loans in the U.S.
Trichet, the European Central Bank (ECB) president at that time, stated that traditional financial theory was not helpful in making policy decisions during the financial crisis and emphasized the need for behavioral economics and multi-agent simulation\cite{Trichet2011}.
Bookstaber, who has worked in risk management at investment banks and hedge funds and has also worked at the U.S. Treasury, argued a paradigm shift to methods that can incorporate complexity, such as multi-agent simulation, in his book\cite{Bookstaber2017}, reflecting on the financial crisis.
It claims that traditional economic theory is difficult to work in times of crisis because distortions are amplified.

There are various studies using artificial market simulation.
Cui et al.\cite{Cui2012} showed that it is impossible to reproduce certain phenomena observed in the stock market only by xero-Intelligence agents.
Torii et al.\cite{Torii2015} conducted a simulation where price shocks spread to other stocks and analyzed the mechanism.
Mizuta et al.\cite{Mizuta2016} analyzed the impact of price quotation (minmum price unit) in the stock market using artificial markets and argued that lowering price quotation is necessary to maintain market share, contributing to discussions on lowering price  quotation at the Tokyo Stock Exchange.
Hirano et al.\cite{Hirano2020c} used artificial markets to analyze the impact of capital adequacy ratio regulation, showing that the capital adequacy ratio regulation, which was supposed to have been introduced to ensure market stability, could amplify price shocks and suppress price increases.
There are also studies that reproduce flash crashes in artificial markets\cite{Leal2019,Paddrik2012}.

Multiple platforms have been proposed to realize these artificial market simulations.
Torii et al.\cite{Torii2017} proposed and released the "Platform for large-scale and high-frequency artificial market" (Plham)\cite{Plham}.
Furthermore, a subsequent platform, the Java version of PlhamJ\cite{Torii2019}, has also been proposed and released.
There are also others such as U-MART \cite{sato2001u}, Santa Fe Artificial Stock Market \cite{arthur2018asset}, and Agent-based Interactive Discrete Event Simulation (ABIDES) \cite{Byrd2020}.
This study presents a new platform, PAMS.

\section{PAMS: Platform for Artificial Market Simulations}
\subsection{Concepts}
The basic concept of PAMS is a user-customizable tick-time scale artificial market simulation that assumes integration with deep learning.
From the perspective of integrating with deep learning, PAMS is released as a Python-based package, and to allow users to easily customize, PAMS adopts an object-oriented class-based architecture, allowing various customizations by overriding.
Moreover, the detailed settings for agents and markets can be set through JSON or Python dictionary easily.

There are time-driven simulations and event-driven simulations in artificial market simulations.
In time-driven simulations, various agents act simultaneously on the same time axis, while in event-driven simulations, time progresses only when an agent takes action.
In artificial market simulations, the former, time-driven simulations, have their challenges.
It is because, in actual financial markets, orders are placed on a microsecond basis, and running a time-driven simulation in actual simulations would require complex calculations at high speed.
Therefore, event-driven simulations based on the placed orders are suitable for artificial market simulations, and PAMS adopts event-driven simulations at tick-time scales.
By adopting tick-time scale simulations, it is possible to reproduce phenomena such as flash crashes where a large number of orders are made in a short time.
Also by thinning the interval between orders, simulations at the level of several years are also possible.

Furthermore, PAMS makes various arrangements so that users can use it easily.
In traditional artificial market simulation platforms such as Plham\cite{Torii2017,Plham} and PlhamJ \cite{Torii2019}, being large-scale was considered important, and many features were incorporated that were thought to assume the use of supercomputers.
For example, Plham is written in X10, a programming language for supercomputers.
However, considering recent improvements in the performance of consumer-oriented computers, it is possible to realize multi-agent simulations even without such large-scale computing resources.
Therefore, PAMS removed the concept of large-scale and set as its development goal the ability to easily execute even in Jupyter Notebook, and developed accordingly.

\subsection{How to Use}
Detailed usage instructions can be found on the Github page\footnote{\url{https://github.com/masanorihirano/pams}} and the official documentation\footnote{\url{https://pams.hirano.dev/en/latest/}}, but here is a brief introduction to how to use it.

First, as shown in Figure \ref{fig:pip}, PAMS can be downloaded from PyPi, the Python package repository.
Of course, it can also be used in pipenv or poetry environments.

\begin{figure}[htb]
  \centering
  \begin{lstlisting}[language=bash,numbers=none,xleftmargin=0pt]
$ pip install pams
  \end{lstlisting}
  \caption{PAMS can be installed via pip command.}
  \label{fig:pip}
\end{figure}

\begin{figure}[htbp]
  \centering
  \begin{lstlisting}[language=Python,xleftmargin=20pt]
import random
import matplotlib.pyplot as plt
from pams.runners import SequentialRunner
from pams.logs import MarketStepSaver
config = {
  "simulation": {
    "markets": ["Market"],
    "agents": ["FCNAgents"],
    "sessions": [
      {	"sessionName": 0,
        "iterationSteps": 100,
        "withOrderPlacement": True,
        "withOrderExecution": False,
        "withPrint": True,
        "hiFrequencySubmitRate": 1.0
      },
      {	"sessionName": 1,
        "iterationSteps": 500,
        "withOrderPlacement": True,
        "withOrderExecution": True,
        "withPrint": True
      }
    ]
  },
  "Market": {
    "class": "Market",
    "tickSize": 0.00001,
    "marketPrice": 300.0
  },
  "FCNAgents": {
    "class": "FCNAgent",
    "numAgents": 100,
    "markets": ["Market"],
    "assetVolume": 50,
    "cashAmount": 10000,
    "fundamentalWeight": {"expon": [1.0]},
    "chartWeight": {"expon": [0.0]},
    "noiseWeight": {"expon": [1.0]},
    "meanReversionTime":{"uniform":[50,100]},
    "noiseScale": 0.001,
    "timeWindowSize": [100, 200],
    "orderMargin": [0.0, 0.1]
  }
}

saver = MarketStepSaver()

runner = SequentialRunner(
    settings=config,
    prng=random.Random(42),
    logger=saver,
)
runner.main()
  \end{lstlisting}
  \caption{An example of simple program. (taken from examples/CI2002.ipynb in Github repository)}
  \label{fig:run-example}
\end{figure}

\begin{figure}[htb]
  \centering
  \includegraphics[width=0.8\linewidth]{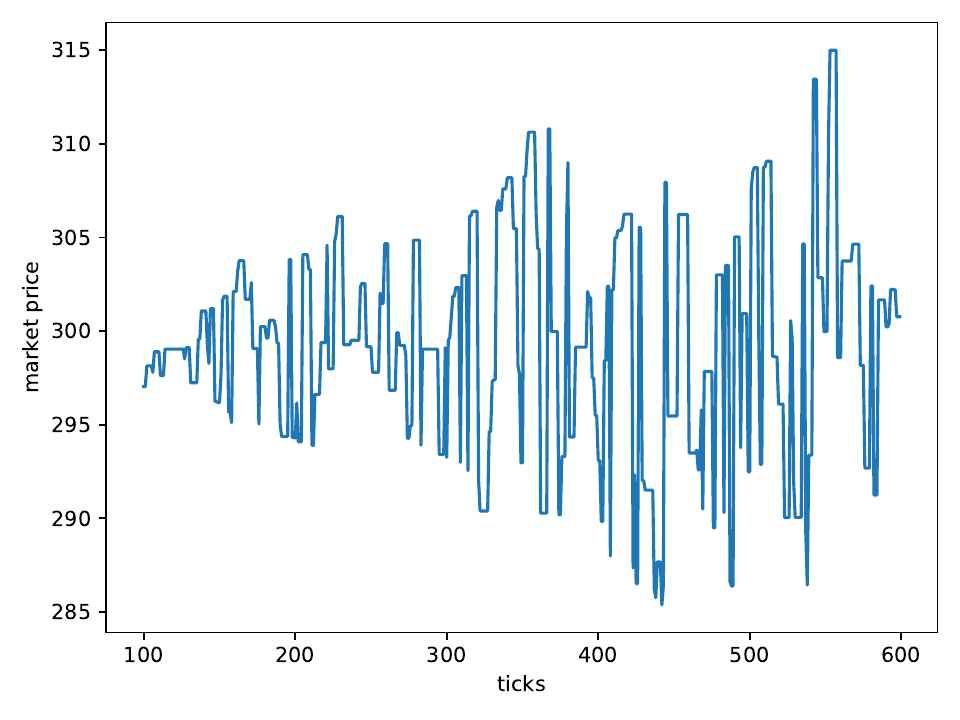}
  \caption{Market prices for each step of the simulation shown in Figure\ref{fig:run-example}.}
  \label{fig:run-price}
\end{figure}

Figure \ref{fig:run-example} is an example of a simple program.
Market and FCNAgents are built-in markets and agents in PAMS, and by using them, you can run simulations easily without complex class overrides.
The config in the figure is the setting of the internal parameters for each market and agent, and these parameters can be set from Python's dictionary type or from an external JSON file.
Also, like Plham, the config setting can accept special entities representing random numbers.
The saver is the instance saving only specific results of the simulation (such as execution and price at each step), and by setting this to runner, it is also possible to calculate necessary statistics after the simulation.
The plot of the price at each step of the simulation using the saver is shown in Figure \ref{fig:run-price}.

\begin{figure}[htb]
  \centering
  \includegraphics[width=\linewidth]{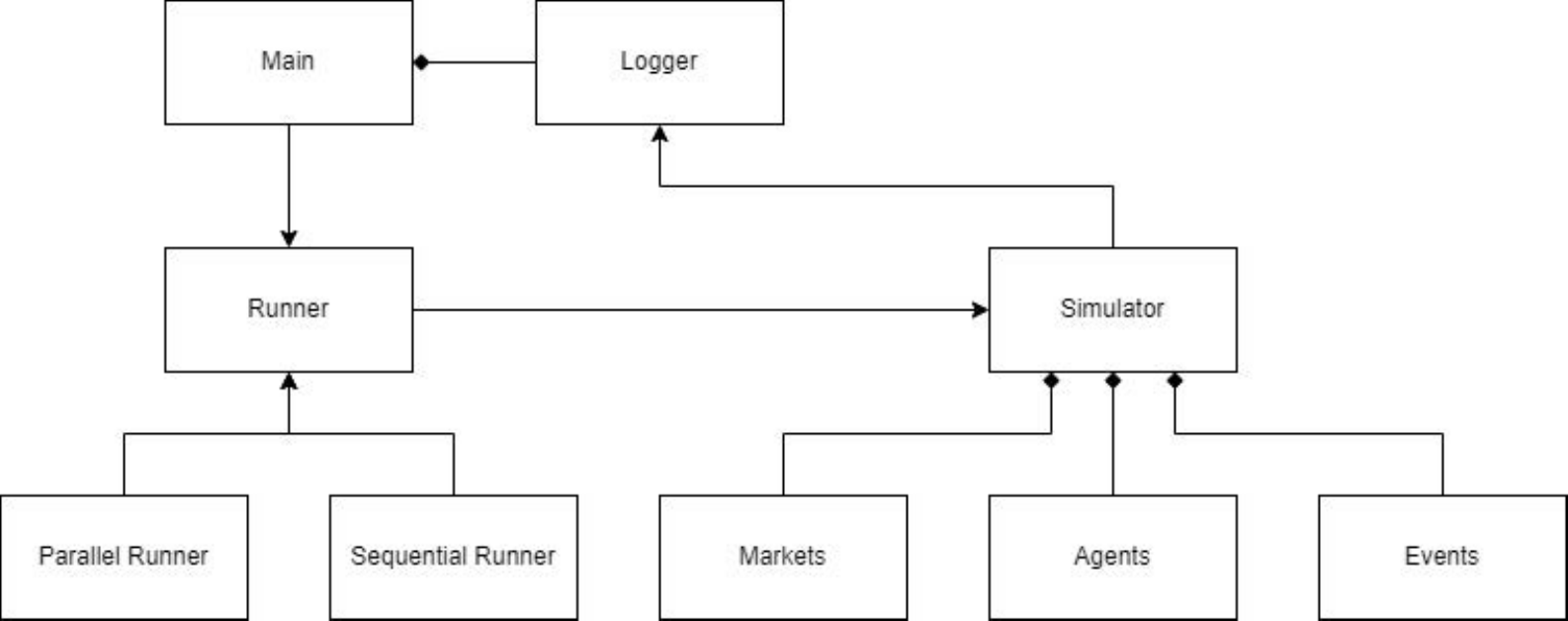}
  \caption{Relationships between abstract classes in PAMS}
  \label{fig:classes}
\end{figure}

More complex customizations can be made than the example shown above.
Figure \ref{fig:classes} shows the relationship between abstract classes.
In PAMS, the Runner class is called from main.
This Runner class controls the order of simulation processing and the calling of agents.
It has a parallel version and a sequential version\footnote{As of July 2023, the parallel version is not implemented}.
This Runner controls the simulations via the interfaces of the Simulator instance.
The Simulator is like a kind of virtual world, where Market, Agent, and Event exist.
In the process, a Logger instance is called and the simulation is recorded.
The Logger instance is defined in the main and provided to the Simulator instance through the Runner instance.

All these abstract classes can be extended, and the Market, Agent, and Event in the Simulator can be easily customized by registering them to the Runner instance.
We have also created examples of using user-defined classes (samples/user\_class in the GitHub repository), so please refer to them as well.

To realize the tick-time scale simulation, at each step of the simulation, a fixed number of agents (default is 1) are randomly called out and given the chance to place orders.

Thus, PAMS realizes a package that can be used universally as a tick-time-based artificial market simulation.

\section{Integration of PAMS and Deep Learning}
In this chapter, we demonstrate an example of integrating PAMS with deep learning.

In agent simulation, deep learning can be used in two ways: in the decision-making process of agents, and in the application of it to simulation output.
In the former case, it is necessary to call the deep learning code inside the simulation, whereas in the latter case, the simulation needs to be called from the deep learning code to obtain data.
In traditional simulations that did not adopt a Python-based architecture, the latter implementation was relatively easy, but the former implementation required calling the deep learning script as a subprocess from the artificial market simulation, which was not easy.

Here, we will tackle the case where a deep learning model is used in the decision-making of agents inside an artificial market simulation, which has been difficult to implement in previous platforms.

\subsection{Task Setting}
In this study, we will investigate the changes in the profit structure of each trader due to the spread of advanced trading strategies.
Here, advanced trading strategies refer to trading based on price prediction using deep learning.
Although there are various methods such as optimizing the trading strategy with reinforcement learning if we use deep learning, for simplicity, we will only focus on the method of trading based on price prediction using deep learning.

In this task, we suppose that there are two types of agents: the general rule-based {\it Stylized Agent} and the {\it Deep Agent} that uses deep learning.
We hypothesize that when deep learning becomes widespread, i.e., when the proportion of Deep Agents increases, they may cancel out the profits of each other, making it difficult to make profits.
We will test this hypothesis.

\subsection{Model}
In this study, we adopt a common artificial market model used in previous research\cite{Torii2015}.
The simulation consists of a continuous double auction market, $n_{\mathrm{sa}}$ Stylized Agents, and $n_{\mathrm{da}}$ Deep Agents.
The total number of agents constant: $n_{\mathrm{sa}}+n_{\mathrm{da}}=100$.
We vary the number of Deep Agents ($n_{\mathrm{da}}$) to see how the profit structure changes.
In following, the details of each agents are explained.

\subsubsection{Stylized Agent}
At time $t$, the stylized trader agent $i$ decides its trading actions using the following criteria: fundamentals, chartists (trends), and noise.
Initially, the agents calculate these three factors.
\begin{itemize}
    \item Fundamental factor:
          \begin{equation}
              F_t^i = \frac{1}{\tau^{*i}} \ln{\left\{\frac{p_t^*}{p_t}\right\}}.
          \end{equation}
          where $\tau^{*i}$ is agent $i$'s mean reversion-time constant, $p_t^*$ is the fundamental price at time $t$, and $p_t$ is the price at time $t$.
    \item Chartist factor:
          \begin{equation}
              C_t^i = \frac{1}{\tau^i}\sum_{j=1}^{\tau^i} r_{(t-j)} = \frac{1}{\tau^i}\sum_{j=1}^{\tau^i}\ln{\frac{p_{(t-j)}}{p_{(t-j-1)}}},
          \end{equation}
          where $\tau^i$ is the time window size of agent $i$ and $r_t$ is the logarithm return at time $t$.
    \item Noise factor:
          \begin{equation}
              N_t^i \sim \mathcal{N} (0, \sigma).
          \end{equation}
          denotes that $N_t^i$ obeys a normal distribution with zero mean and variance $(\sigma)^2$.
\end{itemize}

Subsequently, the agents calculate the weighted averages of these three factors.
\begin{equation}
    \widehat{r_t^i} = \frac{1}{w_F^i + w_C^i + w_N^i} \left(w_F^i F_t^i + w_C^i C_t^i + w_N^i N_t^i\right),
\end{equation}
where $w_F^i, w_C^i$, and $w_N^i$ are the weights of agent $i$ for each factor.

Next, the expected price of agent $i$ is calculated using the following equation:
\begin{equation}
    \widehat{p_t^i} = p_t \exp{\left(\widehat{r_t^i} \tau^i\right)}.
\end{equation}

Subsequently, using a fixed margin of $k^i \in [k_{\mathrm{min}}, k_{\mathrm{max}}]$, we determine the actual order prices using the following rules:
\begin{itemize}
    \item If $\widehat{p_t^i} > p_t$, agent $i$ places a bid (buy order) at the price
          \begin{equation}
              \min{\left\{\widehat{p_t^i} (1-k^i), p_{t}^{\mathrm{bid}}\right\}}.
          \end{equation}
    \item If $\widehat{p_t^i} < p_t$, agent $i$ places an ask (sell order) at the price
          \begin{equation}
              \max{\left\{\widehat{p_t^i} (1+k^i), p_{t}^{\mathrm{ask}}\right\}}.
          \end{equation}
\end{itemize}
Here, $p_{t}^{\mathrm{bid}}$ and $p_{t}^{\mathrm{ask}}$ denote the best bid and ask prices, respectively.

The parameters employed for this type of trader are as follows:
$p_t^*=300, w_F^i \sim Ex(1.0), w_C^i \sim Ex(0.0), w_N^i \sim Ex(1.0), \sigma=0.001, \tau^* \in [50, 100], \tau \in [100, 200], k^i \in [0.0, 0.1]$, which were mainly determined based on \cite{Torii2015}.
Here, $Ex(\lambda)$ indicates an exponential distribution with an expected value of $\lambda$.

\subsubsection{Deep Agent}
The deep learning-based agent predicts the up/down of the price 100 ticks in the future when its turn to act comes, using an LSTM-based neural network based on the past 100 price time series (the price and mid-price series).
For training this prediction model, simulation data before the prediction point is used, and data is created by sliding the time windows.
Of this training data, the last 100 are used as evaluation data for model evaluation.
Only when the prediction accuracy in the evaluation data exceeds 51\%, it is assumed that a good model has been created and is actually used for trading.
If the prediction is up, either place a buy order or, if the position is already 1, leave it as it is.
If the prediction is down, either place a sell order or, if the position is already -1, leave it as it is.

As a neural network architecture, after LSTM processing historical data with a 32-dimensional hidden state, the last output is processed by 2-layered linear layers with Layer Normalization \cite{Ba2016} and ReLU activation, and the logit output for prediction is obtained.

\subsection{Experiment}
As we mentioned earlier, the total number of agents constant: $n_{\mathrm{sa}}+n_{\mathrm{da}}=100$.
We vary the number of Deep Agents ($n_{\mathrm{da}}$) to see how the profit structure changes.
We vary $n_{\mathrm{da}}$ from 1 to 20.

For each situation, we conduct 10 trials and examine the average and variance of the profit of each type of agent.

\subsection{Results}
\begin{figure}[hbt]
	\centering
	\includegraphics[width=\linewidth]{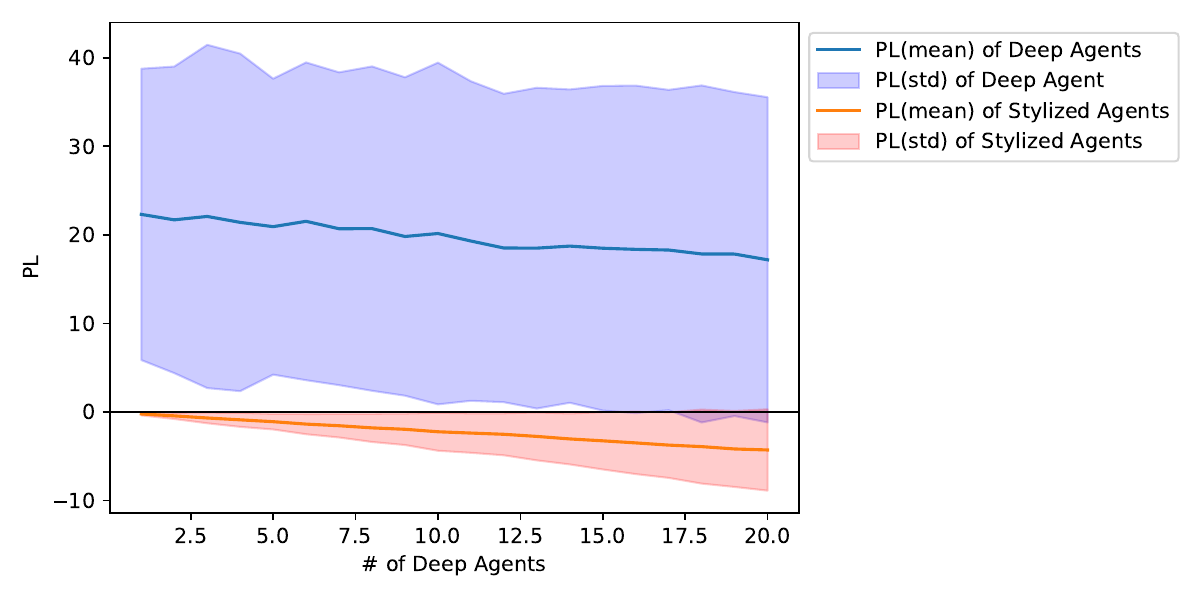}
	\caption{Average profit and loss of Deep Agents and Stylized Agents for varying number of Deep Agents.}
	\label{fig:deep-dl}
\end{figure}
Figure \ref{fig:deep-dl} is a graph of the results.
Looking at this figure, when only Stylized Agents exist, it is almost a zero-sum game.
However, as the number of Deep Agents increases, their average profit becomes negative and smaller.
On the other hand, although Deep Agents have a positive average profit, it seems that the profit of Deep Agents decreases as the number of Deep Agents increases.

\begin{table}[htb]
  \centering
  \caption{Results of regression analysis. *** denotes non-zero at the 99.9\% significance level.}
  \label{tab:my_label}
  \begin{tabular}{@{}lcc@{}}
  \toprule
  Metrics $\backslash$ Agents & Deep Agents & Stylized Agents \\ \midrule
  Intercept & 22.4953*** & -0.0294 \\
  Coefficient & -0.2649*** & -0.2157*** \\ \midrule
  $R^{2}$ & 0.954 & 0.999 \\
  Adjusted $R^{2}$ & 0.952 & 0.999 \\\bottomrule
  \end{tabular}
\end{table}

To analyze these trends more accurately, we conducted a regression analysis, the results of which are shown in Table \ref{tab:my_label}.
Looking at the results, the average profit of Stylized Agents when there are no Deep Agents is not rejected to be zero.
It can be significantly confirmed that the average profit of Deep Agents is positive.
Moreover, as the number of Deep Agents increases, the average profit of both Deep Agents and Stylized Agents decreases.

\subsection{Discussion}
First of all, one thing that can be observed from the experimental results is that Deep Agents can make a profit.
This suggests that even in a simulation environment, Deep Agent can capture appropriate features and trade successfully.
Therefore, the deep learning implemented in Deep Agent can be said to be functioning appropriately as an effective trading strategy.
Moreover, the experiment can be considered appropriate for the task setting.

In addition, the experimental results show that as the number of Deep Agents increases, the average profits of both Stylized Agents and Deep Agents decrease.
It is natural to think that the profits of Stylized Agents will decrease as the number of Deep Agents increases because Deep Agents have a stronger capability to obtain profits than Stylized Agents.
On the other hand, it is not necessarily obvious that the profits of Deep Agents will decrease as their number increases.
For example, if the number of agents with the same trading strategy increases, the number of orders in the same direction may increase, and profits may be amplified.
However, because Deep Agents in this study published a variety of orders, it may have been difficult for price fluctuations with a certain directionality to occur.

According to these results and discussions, the emergence of agents with advanced strategies like Deep Agents not only reduces the profits of existing trading strategies but also cancels out the profits between agents with advanced strategies.

\section{Discussion through the Whole Study}
In this paper, we presented PAMS, a new artificial market simulation platform.
PAMS adopts a Python-based architecture for smooth integration with deep learning, and as demonstrated in this paper, it can easily use deep learning even in the internal structure of agents.
In addition, it allows users to easily customize agents and environments.
No such artificial market simulation platform has existed before.

The potential of PAMS is very broad.
With the ease of integration with simulations and deep learning, it is possible to carry out initiatives to realize more realistic simulations, like a digital twin of the real market, using real data.
In addition, it will now be easier to create agents using deep reinforcement learning, not just a simple deep learning-based trading strategy as shown in this paper.
Also, because simulation data is also valuable, it may be possible to use simulations as a method of data augmentation.

As a future development, PAMS currently cannot incorporate real-time scales.
This is due to the lack of any trial to incorporate both tick-time scales and real-time scales at the same time in simulations.
However, even in a tick-time scale, if the interval between the previous orders is included in the features of the order, it is possible to have a real-time scale at the same time.
However, it is currently difficult to control this order interval well.
In other words, this order interval is not something that traders themselves can control, especially in the case that markets have high liquidity.
Therefore, careful consideration is needed on how to incorporate the order interval into the model.

Finally, we hope that PAMS will be used in many studies, and further feedbacks that helps us improve PAMS are welcome.
We look forward to many users' trial and error.

\section{Conclusion}
In this paper, we presented PAMS, a new artificial market simulation platform.
PAMS adopts a Python-based architecture for smooth integration with deep learning and other technologies and allows users to easily customize agents and environments.
We also showed an example of a study where deep learning was implemented within a trader agent.
We believe that PAMS has great potential as a platform, and we look forward to its increased use.

\bibliographystyle{IEEEtran}
\bibliography{cite}

\end{document}